\documentclass[letter]{aa}
\usepackage{txfonts}
\usepackage{graphicx}
\begin{document}
   \title{Considerations for the habitable zone of super-Earth planets in Gliese 581} 
   \subtitle{ }

   \author{P. Chylek \inst{1}\and
          M. R. P\'erez \inst{2}
          }

   \offprints{P. Chylek}

   \institute{Los Alamos National Laboratory, P.O. Box 1663, Space and Remote Sensing, MS B244, Los Alamos, NM 87545, U.S.A. \\
              \email{chylek@lanl.gov}
         \and
             Los Alamos National Laboratory, P.O. Box 1663, Space Sciences and Applications, MS B244, Los Alamos, NM 87545, U.S.A. \\
		\email{mperez@lanl.gov}
             }

   \date{Received September 2007/ Accepted XXXX 2007}

	\abstract{}{We assess the possibility that planets Gliese 581 c and d are within the habitable zone.}{In analogy with our solar system, we use an empirical definition of the habitable zone. We include assumptions such as planetary climates, and atmospheric circulation on gravitationally locked synchronous rotation.}{Based on the different scenarios, we argue that both planets in Gliese 581 could develop conditions for a habitable zone.}{In an Earth-like environment planet d could be within a habitable zone, if an atmosphere producing greenhouse effect of $100K$ could have developed. If the planets are gravitationally locked-in, planet c could develop atmospheric circulation that would allow it to reach temperatures consistent with the existence of surface liquid water, which in turn could support life.}
\keywords{stars: individual Gliese 581 -- stars: planetary systems – super-Earth planets – astrobiology}
\maketitle
%
\section{Introduction}
M stars are the most common stars in our galactic vicinity and about 75\% (Tarter et al. 2006) of all stars in the Galaxy. Because of a lower mass (compared to our Sun) the M stars radiative output, after an initial highly variable stormy period, becomes stable for many billions of years, raising a possibility of evolution of life on some of their planets. Following the discovery of a Neptune-mass planet around Gliese 581 (G581 hereafter), planet b, by Bonfils et al. (2005), Udry et al. (2007) discovered two additional super-Earth planets, G581 c and d, and, based on solely orbital considerations, determined that planet c was within the habitable zone~(HZ).

Later, adding atmospheric considerations or greenhouse effects, von Bloh et al. (2007a) recently concluded that planet c is outside the HZ, whereas planet d is well within the HZ.  

G581 has a mass of about 0.31M$_{\sun}$, spectral type M3 dwarf, and its photospheric temperature is estimated to be T$_{eff}=3480K$, compared to the T$_{eff}$ of $5770K$ for the Sun. G581 luminosity has been estimated to be $L_{G581}=0.013\pm 0.002 L_{\sun}$. The planets' masses and their distances (semi-major axis) from the central star are M$_{c}=5.06M_{\oplus}$, R$_{c}=0.073~AU$, and M$_{d}=8.3M_{\oplus}$ with R$_{d}=0.25~AU$, respectively. 

In this Letter, we consider the physical conditions of planets c and d to be similar to conditions of the inner planets in the solar system and we show that under special, but feasible circumstances, both planets could have developed HZs.
\section{Limits on the solar system HZ}   
HZ is generally defined as a zone within which a planet can maintain liquid water on its surface. Different sets of assumptions have been used in the published literature to specify the inner and the outer boundary of a HZ. Instead of considering advantages and disadvantages of various definitions we use empirical boundary limits deduced from the observations of our own solar system. We know that the Earth is within the HZ of our Sun, while Venus is outside the inner boundary, and Mars is outside the outer boundary.  

At some point in early stages of planetary evolution all three considered planets (Venus, Earth and Mars) were stripped of their original hydrogen rich atmospheres. The secondary atmospheres were built up by releasing mostly water vapor and carbon dioxide from planetary interiors. Due to the differences in planetary effective surface temperature, gravity, kinematics and geological composition, the atmospheres of the three considered planets, evolved into the current diverse environments. 

Although Venus, Earth and Mars have very different surface albedos today (0.6 for Venus, 0.3 for Earth and 0.15 for Mars) we can envision that at some point in the past, after the original atmosphere was lost and before the secondary atmosphere was formed by interior out-gassing, all three planets had approximately the same surface albedo given basically by surface reflectivity of the rocky surface material. We assume that the surface albedo at that time was around 0.1, somewhere in between the current albedo of Mars ($a=0.15$) and Mercury ($a=0.05$). To obtain effective planetary surface temperatures at the time after the planetary primitive atmosphere was lost and before the secondary atmosphere was formed, we use a simple expression for an effective planetary temperature, T$_{eff}$, implied by a top-of-atmosphere radiative balance,
$$ T_{eff}=[{\frac{S_o(1-a)}{4\sigma}}]^{1/4} \eqno(1)$$                                                                                              
\noindent
where, $S_o$ is the incident solar flux at considered location and time, $a$ is planetary albedo, and $\sigma$ the Stefan-Boltzman constant, and we assume the planet to be a black body radiator ($\epsilon_1=1$) and a gray absorber ($\epsilon_2=1-a$).  We also assume the solar luminosity to be reduced by 26\% (Harwit 2006 - for about 4.5 billion years ago) and we calculate the $T_{eff}$ as a function of the planetary albedo. Considering the surface albedo to be somewhere in between 0.05 and 0.15 (current surface albedos of Mercury and Mars), we detect in Fig. 1: top panel, that the effective temperatures were around $T_{eff}(Mars)=200K$, $T_{eff}(Earth)=250K$ and $T_{eff}(Venus)=290K$. 
\begin{figure}
   \centering
 \includegraphics[angle=0,width=7.5cm]{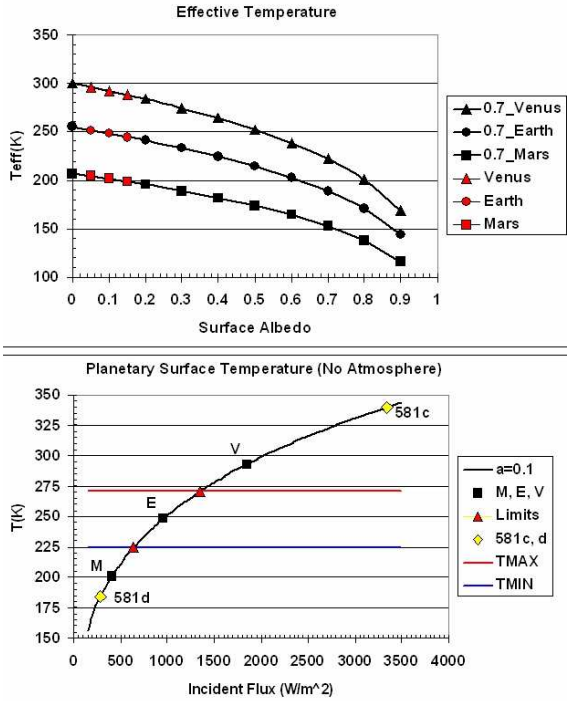}
      \caption{Top panel: Effective temperatures of Venus (V), Earth (E) and Mars (M) as a function of planetary albedo. Bottom panel: Effective temperatures of ``naked'' V, E, M, planets c and d after the loss of their primitive atmosphere and before acquiring their secondary atmosphere, and empirical boundaries of a HZ (horizontal red and blue lines). }
         \label{Fig1}
   \end{figure}
From empirical observations we know that of three planets only Earth has made it into the HZ. Although it is difficult to determine the inner and the outer boundaries of the HZ, we take the half points between Earth and the other planet's (Venus for the inner and Mars for the outer boundary) effective temperature as the boundary of Earth's HZ (Fig. 1: bottom panel). We consider all three planets to have at the time a surface albedo of 0.1, and we assign half of the HZ temperature width as the uncertainty of our estimate. Thus we obtain the lower temperature limit of the HZ at the time before formation of the secondary planetary atmosphere to be $T_{MIN}=225\pm14~K$ and $T_{MAX}=270\pm11~K$ for the upper limit. Translating the HZ boundary within our solar system into the distance from the Sun, we determine $R_{MIN}=0.82\pm0.09~AU$ and $R_{MAX}=1.19\pm0.09~AU$.

\section{G581 c and d planets}
Now we consider the question of whether it is possible for planets c and d to develop into a stage in which they can keep liquid water on their surface. We assume that the primitive original atmospheres were swept away by stellar flares during an early evolution of G581 and we further assume that its luminosity did not change significantly after its arrival on the main sequence. After the original primitive atmosphere of the planets was lost and before a possible secondary atmosphere could be built there was a time when the planets were without any atmosphere. The albedo of these planets was solely determined by reflectivity of its surface rocky layer. We, therefore, assume the planetary surface albedo of 0.1 and calculate the effective surface temperature of the planets c and d during the time after they lost their primitive atmospheres and before the formation of the secondary atmospheres by out-gassing from the planets' interiors. 

The results (Fig. 1: bottom panel) indicate that both planets are outside the range of temperature that would allow a HZ with a possibility of liquid water on the surface.  It is apparent that the temperature of the planet c is too high (higher than temperature of Venus before the runaway greenhouse effect) to develop a habitable environment and the temperature of planet d is too cold (colder than the temperature of Mars). 

We suggest that under conditions similar to our solar system, planet c would proceed on a path similar to Venus and would become too hot to allow a liquid water to exist on its surface.  On the other hand planet d would follow a path similar to Mars and would stay too cold. We note that during the considered situation, namely before the formation of the secondary atmosphere, planet c was warmer than Venus by at least $35K$ and planet d was colder than Mars by at least $15K$. 

We conclude that neither of these planets could have developed conditions allowing a liquid water to exist. This conclusion is subject to the assumptions that the processes involved in the evolution of the secondary atmospheres proceeded on these planets in a fashion similar to the processes that took place in our solar system within the Venus, Earth and Mars environment. Next we consider conditions that might have been sufficiently different from conditions in the solar system to allow a colder environment for planet c and a warmer environment on planet d. 

\section{Planetary climate}
The secondary atmosphere, produced by out-gassing from the planetary interior, is composed mainly of water vapor and carbon dioxide. In the case of a suitable Earth-like environment, most of the water vapor condenses to form oceans and ice caps with only a minor part staying in the atmosphere. Carbon dioxide becomes bound to rocks and absorbed by oceans and the biosphere, again leaving only a minor portion in the atmosphere. 

The formation of a secondary atmosphere has two major means to affect the surface temperature. The first one is through the ``greenhouse effect'' which essentially reduced the flux of thermal infrared radiation to space. The second mode of surface temperature modification is through the planetary albedo changes connected predominantly to the cloud formation. In the case of Earth, the planetary albedo is decreased (compared to the bare rocky surface) by the presence of oceans, and increased by the clouds in such a way that the average planetary albedo changes from its estimated original ``rocky surface'' value of 0.1 to the current Earth planetary albedo of about 0.30. The albedo change by itself would decrease the effective radiative temperature of the Earth by about $16K$, from the value of $271K$ (for the current value of solar flux and albedo of 0.10) to $255K$ (the current value of solar flux and albedo 0.30).

However, the water vapor and carbon dioxide greenhouse effect contributes to the warming of about $33K$ and increase the average of surface temperature to about $288K$. About a half of greenhouse warming is utilized to compensate for increase of albedo (from 0.1 to 0.3 and connected temperature decrease of $16K$) and the other half produces an increase of the surface temperature, by about $17K$, from the original $271K$ to $288K$. 

If the temperature of a planet is too low, the water vapor and carbon dioxide will condense into ice and will be removed from the atmosphere and deposited on the surface. The result is an increased planetary albedo and an atmosphere lacking any greenhouse effect (Mars-like environment). In the opposite case of too high a planetary temperature, all water vapor and carbon dioxide stay in the atmosphere. The planetary albedo will significantly increase due to a high reflectivity of water or $CO_2$ clouds, however, the strong greenhouse effect will dominate and compensate for an increased albedo and cause an additional significant warming of the planet (Venus-like environment).
\begin{figure}
   \centering
 \includegraphics[angle=0,width=7.5cm]{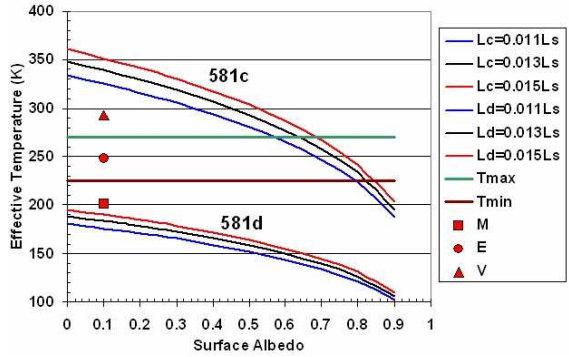}
      \caption{Effective temperatures of ``naked'' Venus (V), Earth (E) and Mars (M), boundary of HZ (horizontal green and red lines), and effective temperatures of planets c and d as a function of their planetary albedo.}
         \label{Fig2}
   \end{figure}
The situation of planets c and d is depicted in Fig. 2. To bring planet c into a HZ, it needs to increase its albedo from the bare planet value of 0.10 to about 0.69 to 0.70. This can hardly be done without covering the whole planet by clouds, which in turn implies a dense $H_2O$ vapor and/or $CO_2$ atmosphere. Such an atmosphere would, however, create a large greenhouse warming that would push the planet back to the region of too high temperature for liquid water to exist on its surface. Thus it seems that under the considered scenario the planet c has a little chance of developing conditions for liquid water to exist on its surface.

The bare planet d is too cold to hold liquid water ($T_{eff}=183K$). We can assume that a medium strength (comparable to Earth) atmospheric greenhouse effect could provide about $35K$ warming. The accompanying cloudiness would increase the planetary albedo to about 0.30, which would produce a cooling of about $11K$. Thus the result of Earth-like atmosphere would produce a net surface warming of about $24K$, which is not sufficient to bring the planet into the range of habitability. However, a considerably denser atmosphere that would lead to a planet completely covered by clouds, producing a planetary albedo of about 0.6, and warming of around $100K$ (about three times the greenhouse effect of the Earth's atmosphere) would be sufficient to bring the planet d into a HZ. No objectionable mechanism is apparent that would prohibit the formation of such an atmosphere. 

We conclude that assuming a situation similar to our solar system, it is very unlikely that the planet c could be within a HZ. Whereas, the planet d with an atmosphere producing three times the greenhouse effect on Earth, could be within the HZ.

\section{Atmospheric circulation on a gravitationally locked planet}
Climate and atmospheric circulation on gravitationally locked planets is not currently well understood. For the case of G581 planets, based on kinematics considerations, von Bloh et al. (2007a) invoked the notion that planets inside the HZ of M stars are tidally locked. To better understand and justify this assumption we tested this hypothesis in our solar system planets. 

We define the dimensionless parameter, $\phi$, as the ratio between the planet orbital period (in terrestrial days) to the length of the planetary day (in terrestrial hours). Obviously, for tidally locked planets, $\phi$ is equal to 0. In Table 1, we present the values of $\phi$ for the eight recognized planets noting that at the distance of Mercury from the Sun, $d=0.39~AU$, $\phi$ value is small suggesting that Mercury is close to being a tidally locked planet. Consequently, at the estimated distances of planets c and d from the central star G581, of $0.073~AU$ and $0.25~AU$, respectively, the assumption of gravitationally locked planets seems reasonable when compared with our solar system. We also included in Table 1 for G581 and solar system planets the gravitational strength values demonstrating that planets c and d are considerably more tidally locked than any planet in our solar system.   
\begin{table*}
\caption[]{Solar system values of $\phi$ and the value of the pull or gravitational strength $(kg^2/m^2)$ relative to Earth.}
\tabcolsep0.11cm
\begin{center}
\begin{tabular}{cccccccccccc}
\hline\noalign{\smallskip}
&G581 c&G581 d&Mercury&Venus&Earth&Mars&Jupiter&Saturn&Uranus&Neptune\\
\noalign{\smallskip}
\hline\noalign{\smallskip}
$\phi$&--&--&0.02&0.08&15.22&27.8&437.5&1004.4&1778.4&3714.3 \\
$\frac{M_p\times M_s}{R^2}$&294.35&41.17&0.37&1.56&1.00&0.05&11.74&1.04&0.04&0.02 \\
\noalign{\smallskip}
\hline
\end{tabular}
\end{center}
\end{table*}
 
Now we present basic qualitative ideas and we emphasize the need to modify existing atmospheric general circulation models to investigate the problem in detail. Since only a half of a planet is exposed to incident solar radiation, the radiative energy balance equation needs to be modified to reflect this situation. We have, under similar assumptions as in (1),
$$  S_o(1-a)\pi r^{2} = 2 \pi r^{2} \sigma (T^{4} + T_{o}^{4}), \eqno(2) $$
\noindent                                                                     
where, $S_o$ is an incident radiative flux, $a$ is an albedo of an illuminated part of a planet, $r$ is planet's radius and $T$ and $T_o$ are the effective temperatures of an illuminated and shadow side, respectively. If there is no energy transport between the illuminated and shadow side of a planet, or if the transport is negligible so that $T_o<<T$ can be neglected we have,
$$  T_{eff}=[{\frac{S_o(1-a)}{2\sigma}}]^{1/4}, \eqno(3) $$                                                                                   
\noindent
and the effective temperature on the illuminated side increases by about 19\% compared to the case of planet that is not gravitationally locked-in. At the same time the temperature on the shadow side decreases and without atmospheric circulation and possible heat conduction through the planetary interior it would be close to the cosmic background temperature. However, the differential heating of the illuminated side will produce atmospheric circulation, which will initiate a heat transport between the illuminated and not illuminated hemisphere. The maximum rate of heating will occur at the center of the illuminated side of the planet. Here the air rises up creating a low pressure region near the surface of the central region of the illuminated side (Fig. 3). We acknowledge that serious climate stability considerations are problematic under this scenario (Porto de Mello, del Peloso \& Ghezzi 2006).   

Due to the axial symmetry of the illumination the surface air from all directions will move towards the low pressure at the center of the illuminated side while the upper air will diverge from the center into all directions towards the edges of the shadow side. The upper air will experience radiative cooling as it moves from the center towards the edges and, when sufficiently cool, will start descending towards the surface. This thermally driven circulation is an analogy to the Hadley circulation cell of the Earth atmosphere. The differences are, however, quite significant (e.g, the absence of a Coriollis force and an axial symmetry of circulation on a gravitationally locked-in planet). 

In this way, two distinct regions of the interaction between the air of the illuminated and shadow sides can develop. Still partially warm upper air moving away from the central region of an illuminated side can ``overshoot'' and enter the shadow part of the planet. The low level cold surface air from the shadowed side can flow into the lower pressure regions towards the center part of the illuminated side. How much of these possible air exchanges really takes place will depend on the parameters of the locked-in planet atmosphere that enters into the Navier-Stokes equations of motion. 

Qualitatively the atmospheric circulation will try to decrease the temperature gradient between illuminated and shadow parts of the planet. To assess how much heat could be transported if planets c and d are gravitationally locked-in will require detailed knowledge of the atmospheric properties of the planets. A simple example based on an energy balance indicates that to keep the shadowed part at the temperature of $200K$ and the illuminated part at $300K$ would require about 20\% $((T_o/T)^4\sim 0.20)$ of the incident energy to be transported towards the shadow side of the planet, while $250K$ and $300K$ temperatures would require almost 50\% $((T_o/T)^4\sim 0.48)$ of incident energy to be transported to the shadowed side. Thus it would be extremely difficult to maintain on the planet d the shadow side temperature at the level that would prevent water vapor and carbon dioxide from being removed from the atmosphere. 
\begin{figure}
   \centering
 \includegraphics[angle=0,width=7.5cm]{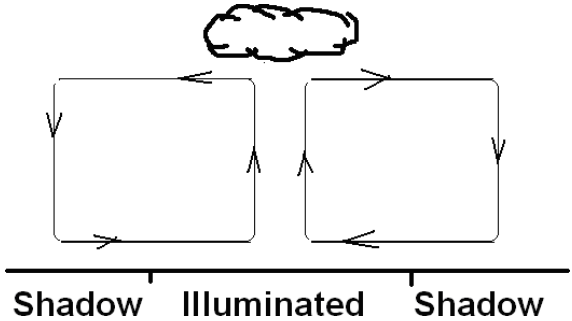}
      \caption{Conceptual illustration of atmospheric circulation on a gravitationally locked-in planet.}
         \label{Fig3}
   \end{figure}
On the other hand planet c, in the locked-in situation may have a chance to develop a habitable region. To keep the shadowed site of the planet at the temperature of about $250K$ would require about 26\% of the heat to be transported to the shadowed side, while about 15\% transport would be sufficient to maintain the average temperature at $220K$. Considering a possible greenhouse effect of about double that found currently in Earth's atmosphere, we can reach a surface temperature around $286K$, just right to almost duplicate our earthly environment. Of course, the problem would be the overheated illuminated side of the planet at about $350K$ and a lack of sunshine on the shadow side. Perhaps tropical forests on the hot side can provide a sufficient amount of oxygen for the whole planet and near the illuminated/shadowed boundary a habitable region may occur.  
    
\section{Conclusion}
We use an empirically defined boundary of a HZ in analogy with our solar system and the fact that only the Earth is within the HZ, while Mars and Venus are not.  By assuming the formation of the secondary atmosphere similar to Earth, we find that neither planet, c nor d, would be within a HZ. However, for planet d there is a possibility to form a HZ provided that a greenhouse effect of about $100K$ could be developed. Conversely, planet c seems to be too hot and we are unable to foresee any situation under which a region of habitability could be developed, under Earth-like conditions. 

If the planets are gravitationally locked-in, as we have shown to be a reasonable assumption, there seems to be a possibility especially for planet c to develop atmospheric circulation that would transport heat from the illuminated to the shadow part of the planet and form a definite region on the planet where liquid water could exist. The future assessment of this possibility will require the development of a simplified Budyko-style climate model (Budyko 1969, Chylek and Coakley 1975), as well as modification of atmospheric general circulation models and additional information concerning the planets and their atmospheric composition. 
 
We found that a gravitationally locked-in planet can have liquid water on its surface even if it is outside a HZ, defined traditionally by the parameters and distance to the sustaining star (Udry et al. 2007). Since half of the planet is permanently in ``sunshine'' an atmospheric circulation can develop between the illuminated and shadow parts and create a HZ with a possibility of surface liquid water in a planet that, otherwise, would be unsuitable. Thus the auxiliary definition of a HZ should be extended to include the rotational and atmospheric characteristics of planets; a concept alluded to by von Bloh et al. (2007a, b).      

\begin{acknowledgements}
The reported research (LA-UR-07-5866) was partially supported by Los Alamos National Laboratory Directed Research and Development Project (20050014DR).
\end{acknowledgements}

\end{document}